\documentclass{article}
\usepackage{emulateapj,psfig}
\usepackage{graphicx}
\usepackage{timesfonts}
\makeatletter
\newenvironment{inlinetable}{%
\def\@captype{table}%
\noindent\begin{minipage}{0.999\linewidth}\begin{center}}
{\end{center}\end{minipage}\smallskip}

\newenvironment{inlinefigure}{%
\def\@captype{figure}%
\noindent\begin{minipage}{0.999\linewidth}\begin{center}}
{\end{center}\end{minipage}\smallskip}
\makeatother
\def\ltsima{$\; \buildrel < \over \sim \;$}
\def\lsim{\lower.5ex\hbox{\ltsima}}
\def\loe{\lower.5ex\hbox{\ltsima}}
\def\gtsima{$\; \buildrel > \over \sim \;$}
\def\gsim{\lower.5ex\hbox{\gtsima}}
\def\goe{\lower.5ex\hbox{\gtsima}}

\def\deg {^\circ}

\newcommand{\be} {\begin{equation}}

\newcommand{\ee} {\end{equation}}
\newcommand{\etal}{ et al. }
\newcommand{\src}{1RXS\,J170849.0--400910}
\newcommand{\R}{{\em ROSAT}}

\newcommand{\bc}{\begin{center}}
\newcommand{\ec}{\end{center}}
\newcommand {\rc}{\rm}
\lefthead{ISRAEL ET AL.}
\righthead{FURTHER EVIDENCE THAT \src\ IS AN ANOMALOUS X--RAY PULSAR}
\submitted{Received: 25 February 1999;~~ accepted: 15 April 1999}

\begin{document}

\title{Further evidence that \src\ is an Anomalous X--ray pulsar}

\authoremail{gianluca@coma.mporzio.astro.it}

\author{G.L. Israel\altaffilmark{1,2}, S. Covino\altaffilmark{3}, 
L. Stella\altaffilmark{1,2}, S. Campana\altaffilmark{3,2}, 
S. Mereghetti\altaffilmark{4,5} }

\affil{1. Osservatorio Astronomico di Roma, Via Frascati 33,
I--00040 Monteporzio Catone (Roma), Italy; Gianluca.Israel and 
Luigi.Stella@oar.mporzio.astro.it}
\affil{2. Affiliated to the International Center for Relativistic
Astrophysics (I.C.R.A.)}
\affil{3. Osservatorio Astronomico di Brera, Via Bianchi 46, I--23807
Merate (Lc), Italy; covino and campana@antares.merate.mi.astro.it}
\affil{4. Istituto di Fisica Cosmica ``G.P.S. Occhialini'' del C.N.R., 
Via Bassini 15, I--20133 Milano, Italy; sandro@ifctr.mi.cnr.it}
\altaffiltext{5}{The results reported in this Letter are partially 
based on observations carried out at ESO, La Silla, Chile (61.D--0513).}

\begin{abstract}
We report the results of two \R\ HRI observations of  
the recently discovered $\sim$11\,s X--ray pulsar \src. A refined position 
with a smaller error radius 
(10\arcsec\ uncertainty) and a new spin period measurement were obtained. 
These results   allowed to derive a period derivative of $\sim$ 7 $\times$ 
10$^{-4}$ s yr$^{-1}$ and to perform a  photometric and 
spectroscopic study of the possible optical counterparts of the source. 
The limits derived from the optical to X--ray flux ratio exclude
the presence of a massive OB companion.

These findings, together with
the nearly constant X--ray flux across observations spanning
three years,
strongly support the inclusion of this $\sim$11 s pulsar in the class
of Anomalous X--ray Pulsars (AXPs).

\end{abstract}

\keywords{stars: neutron --- pulsar: individual (\src) 
--- pulsars: general --- X--ray: stars}

\section{Introduction} 
The  relatively bright X--ray source \src\ was originally discovered 
in the \R\ All Sky Survey (RASS, Voges \etal 1998), but it did not 
attract much attention until $\sim$11\,s periodic pulsations were found
during an ASCA observation that was part of the Galactic Plane 
Survey Project (Sugizaki \etal 1997).
Based on the spin period value and the soft X--ray spectrum 
Sugizaki \etal (1997) suggested that this optically unidentified  source 
could be a new member of the small group of AXPs.  
AXPs are a group of X--ray pulsars 
(Mereghetti \& Stella 1995; van Paradijs \etal 1995) that share 
several peculiar 
properties: a narrow interval of pulse periods (6--12 s), very soft X--ray spectra 
(a steep power--law with photon index $\sim$ 3--4 often coupled 
to an additional blackbody--like component with  
kT$\sim$ 0.4--0.6 keV), lower X--ray luminosities ($\lsim$10$^{35}$ 
erg s$^{-1}$) than those of persistent High Mass X--ray Binaries (HMXBs), a 
narrow spatial distribution in the Galactic plane, absence of large 
flux variations on timescales from months to years, spin--down trends at relatively 
stable rates (2--100 $\times$ 10$^{-5}$ s yr$^{-1}$), and absence of massive 
companion stars (see  Mereghetti \etal 1998 for a recent review). 
Whether AXPs are isolated neutron stars or members of binary systems remains to be 
determined; similarly it is unclear whether they are powered by accretion or by magnetic 
energy, as recently suggested by the analogy with Soft $\gamma$--ray Repeaters 
(SGRs; Kouveliotou \etal 1998, Mereghetti 1999). 

In this Letter we report on the data analysis of public \R\ HRI observations 
of \src\  and on a study of its  possible optical counterparts. Our results 
strongly support the suggestion that \src\ is a member of the AXP group.

\section{ROSAT HRI Observation}

The field of \src\ was observed twice by the \R\ High Resolution Imager (0.1--2.4 keV): 
on 1994 March 8 (ROR 900595) for a total exposure time of 1120\,s and on 
1997 March 24--26 (ROR 180188) for a total exposure time of 11175\,s. 
The source was detected at a level of 0.13$\pm$0.01 and 0.130$\pm$0.003 counts 
s$^{-1}$ in the 1994 and 1997 observations, respectively (90\% uncertainties are 
used through out this letter). 
 
We analyzed the two HRI images using both  a sliding cell and a wavelet 
transform--based detection algorithm (Lazzati et al. 1999; Campana et al. 1999).
In the longer  observation (1997) the  source was detected at 
R.A.=17$^h$08$^m$47$^s$.24 and DEC=--40$^o$08\arcmin50\arcsec.7 (J2000), 
with a statistical error  of only 0.15\arcsec\ (1$\sigma$).   
However, due to the the uncertainty in the satellite boresight, an overall
error radius of  $\sim$ 10\arcsec\  must be considered. 
The new position is located some 35\arcsec\ and 25\arcsec\ North 
of the ASCA position (30\arcsec\ radius; Sugizaki \etal 1997) and PSPC RASS 
position (9\arcsec\ uncertainty).   
The 1993 \R\ HRI observation was analyzed in the same way and
provided a slightly different (but consistent) position,   
R.A.=17$^h$08$^m$46$^s$.52 and DEC=--40$^o$08\arcmin52\arcsec.5 (J2000; $\sim$ 
10\arcsec\ radius uncertainty; see also Fig.~3).

The \R\ HRI has no intrinsic spectral resolution, therefore in order 
to  obtain the source flux from the observed count rate we
assumed the spectral model derived with ASCA (Sugizaki 
\etal 1997), i.e. a power--law with photon index $\Gamma$=3.45 and 
N$_H$=1.8$\times$10$^{22}$ cm$^{-2}$. In this way we 
determined a flux at the Earth of 1.1$\times$10$^{-11}$ erg cm$^{-2}$ s$^{-1}$
in the 0.1--2.4 keV  energy range.  
For  a distance of 10 kpc, this corresponds to a source unabsorbed  
luminosity of $L_X \sim$ 1.2 $\times$ 10$^{36}$ erg s$^{-1}$   
in the 0.8--10 keV range. 

The event list of \src\ was extracted from a circle of 
$\sim$40\arcsec\ radius (corresponding to an encircled energy of $\sim$90\%) 
around the X--ray position of the 1997 observation. The photon arrival times 
were corrected to the barycentre of the solar system and a  background 
subtracted   light curve was accumulated in   1\,s bins. A power spectrum 
was calculated over the 
entire observation duration. A highly significant peak ($\sim$ 8$\sigma$) was 
found at a frequency of 0.09092712 Hz, corresponding to a period of 10.998\,s.  
To refine the period determination and reduce its uncertainty we adopted 
a phase fitting technique. The best pulse period was determined to be 
P=10.99802$\pm$0.00005 \,s. 
The shape of the modulation is somewhat asymmetric with a pulsed fraction of 
$\sim$38$\pm$4\% (Fig.\,1). {\rc We note that such pulse shape is similar
to that reported by Sugizaki \etal in the 0.8--2 keV
range (1997; see their Fig.\,3). }
Owing to poor statistics the periodicity was not detected during the 1994 
observation {\rc (assuming a 3$\sigma$ detection threshold)}.  
%\begin{figure*}[!tbh]
%\psfig{figure=RXS1708pul_apje.ps,width=8cm,height=5.cm} 
%\caption{The 1997 March 24--26 \R\ HRI   light curve folded at 
%the best period  P=10.99802\,s.}
%\end{figure*}
\begin{inlinefigure}
\bigskip
\centerline{\includegraphics[width=0.53\linewidth]{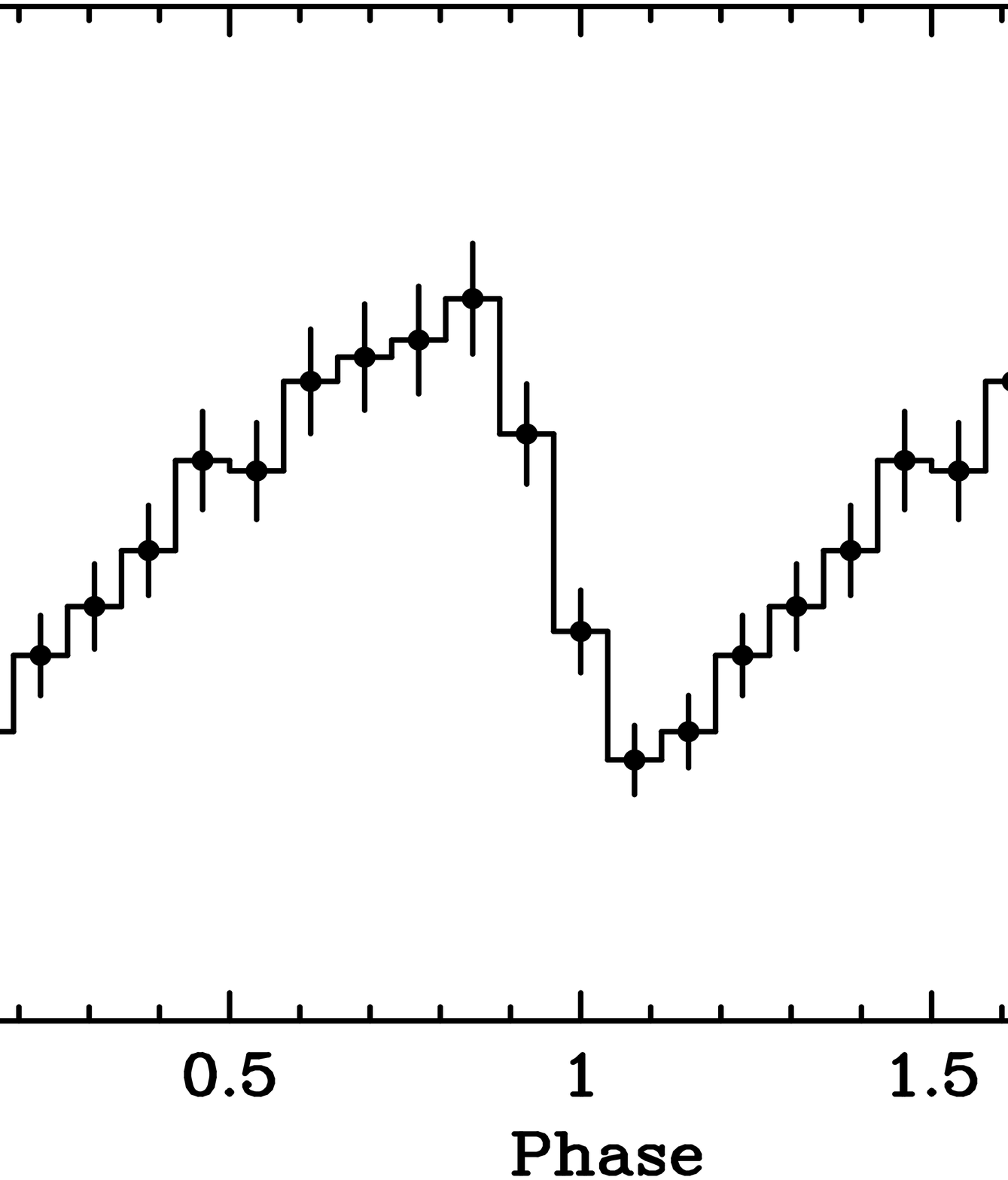}}
\caption{The 1997 March 24--26 \R\ HRI   light curve folded at 
the best period  P=10.99802\,s.}
\bigskip
\end{inlinefigure}
%\begin{figure*}[!tbh]
%\psfig{figure=RXS1708_hr_apje.ps,width=6cm,height=7cm} 
%\caption{Color--Magnitude Diagram   for the objects in the 
%field of \src. The Main Sequence lines for 
%different values of the distance and   reddening are also shown
%(700\,pc, A$_{\rm V}$=1.6; 3 kpc, A$_{\rm V}$=6.2). 
%The straight line  indicates the direction of the reddening  
%and corresponds to a value of A$_{\rm V}$=2. 
%The squares mark the objects within the error boxes.}
%\end{figure*}
\begin{inlinefigure}
\bigskip
\centerline{\includegraphics[width=0.64\linewidth]{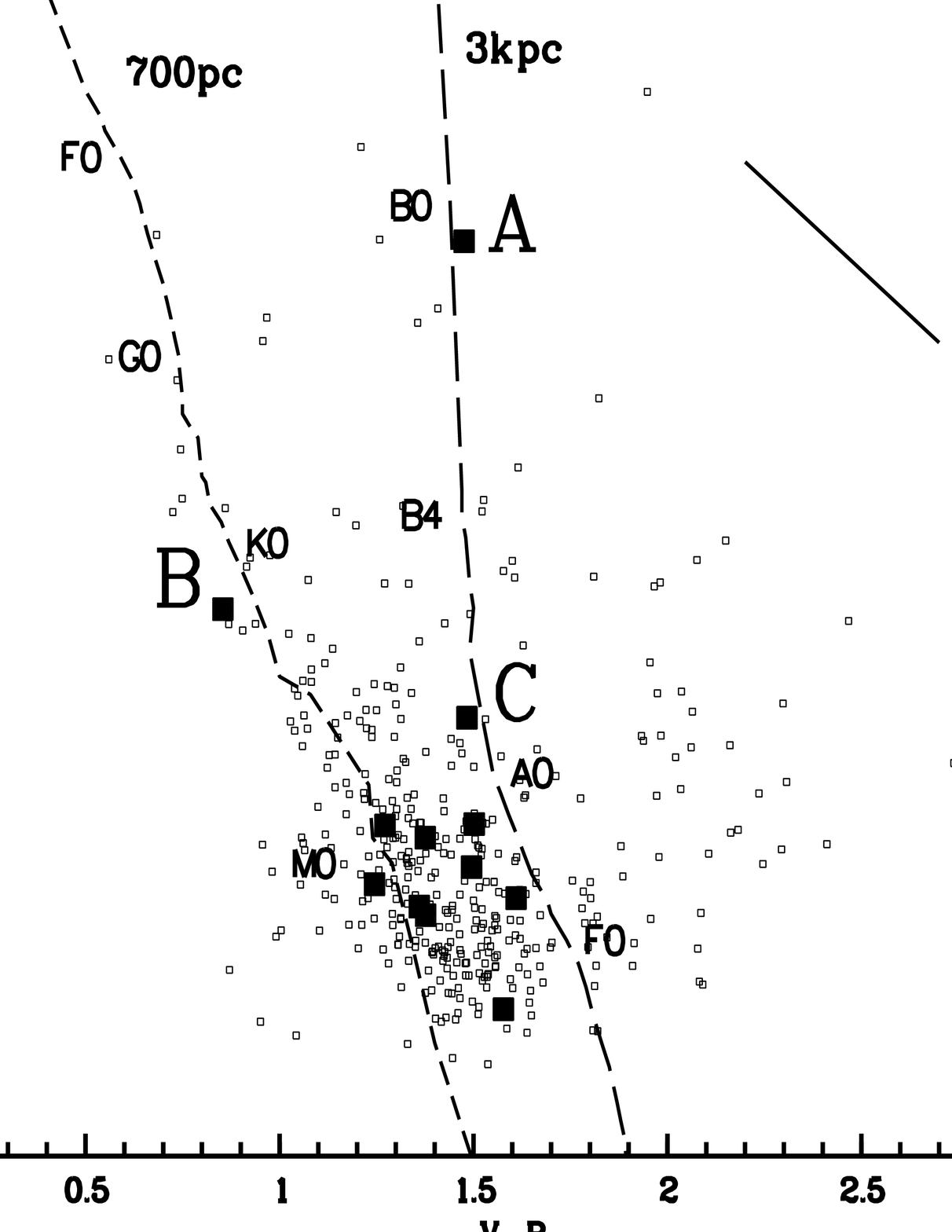}}
\bigskip
\caption{Color--Magnitude Diagram   for the objects in the 
field of \src. The Main Sequence lines for 
different values of the distance and   reddening are also shown
(700\,pc, A$_{\rm V}$=1.6; 3 kpc, A$_{\rm V}$=6.2). 
The straight line  indicates the direction of the reddening  
and corresponds to a value of A$_{\rm V}$=2. 
The squares mark the objects within the error circles.}
\end{inlinefigure}

By comparing our period value with the one measured with  ASCA on 1996 September 3  
(Sugizaki \etal 1997) a period derivative \.P = (7$\pm$2)$\times$10$^{-4}$ s yr$^{-1}$ 
was inferred. \\

\section{Optical Imaging and Spectroscopy} 

Photometry for each stellar object in the images was derived by means of DAOPHOT\,II 
(Stetson 1987), and a Color--Magnitude Diagram (CMD) was then computed (see Fig.\,2) 
for all stars within a region of 4\arcmin$\times$5\arcmin\ around the 
X--ray position of \src.  
For comparison Fig.\,2 also shows the main sequence
for two representative distances: 700\,pc (with A$_{\rm V}$=1.6) and 3\,kpc
(A$_{\rm V}$=6.2). 
Note that \src\ is located in the Galactic plane (l$^{II}$=346.48, b$^{II}$=0.03) 
and that along  this direction spiral arms are located at a 
distance of $\sim$\,1kpc, $\sim$ 3\,kpc and $\sim$4.5\,kpc (Taylor \& Cordes 1993).
The R image of the field of \src\ is shown in  Fig.\,3. The new X--ray position 
uncertainty regions  obtained with the two HRI observations are shown (10$\arcsec$ 
radius). 
Table\,1 gives  the positions, R magnitudes and the colors of the   
objects within (or close to) the error circles. 

The spectrum of the brightest object within the error circle  
(candidate  A, R = 13.4; V--R = 1.48) is shown in Fig.\,4. 
The object was classified as a B0V spectral--type star with  an absorption 
A$_{\rm V}\sim$6.2 . We found no evidence for any emission--lines  in its spectrum.      
%\begin{figure*}[!tbh]
%\psfig{figure=RXS1708_R_apje.ps,width=8cm,height=7.8cm} 
%\caption{R image of the field of \src. 
%The new X--ray position uncertainty regions (10\arcsec\ radius) obtained 
%with the 1994 and 1997 HRI observations are also shown. 
%North is top, east is left.}
%\end{figure*}
\begin{inlinefigure}
\bigskip
\centerline{\includegraphics[width=0.83\linewidth]{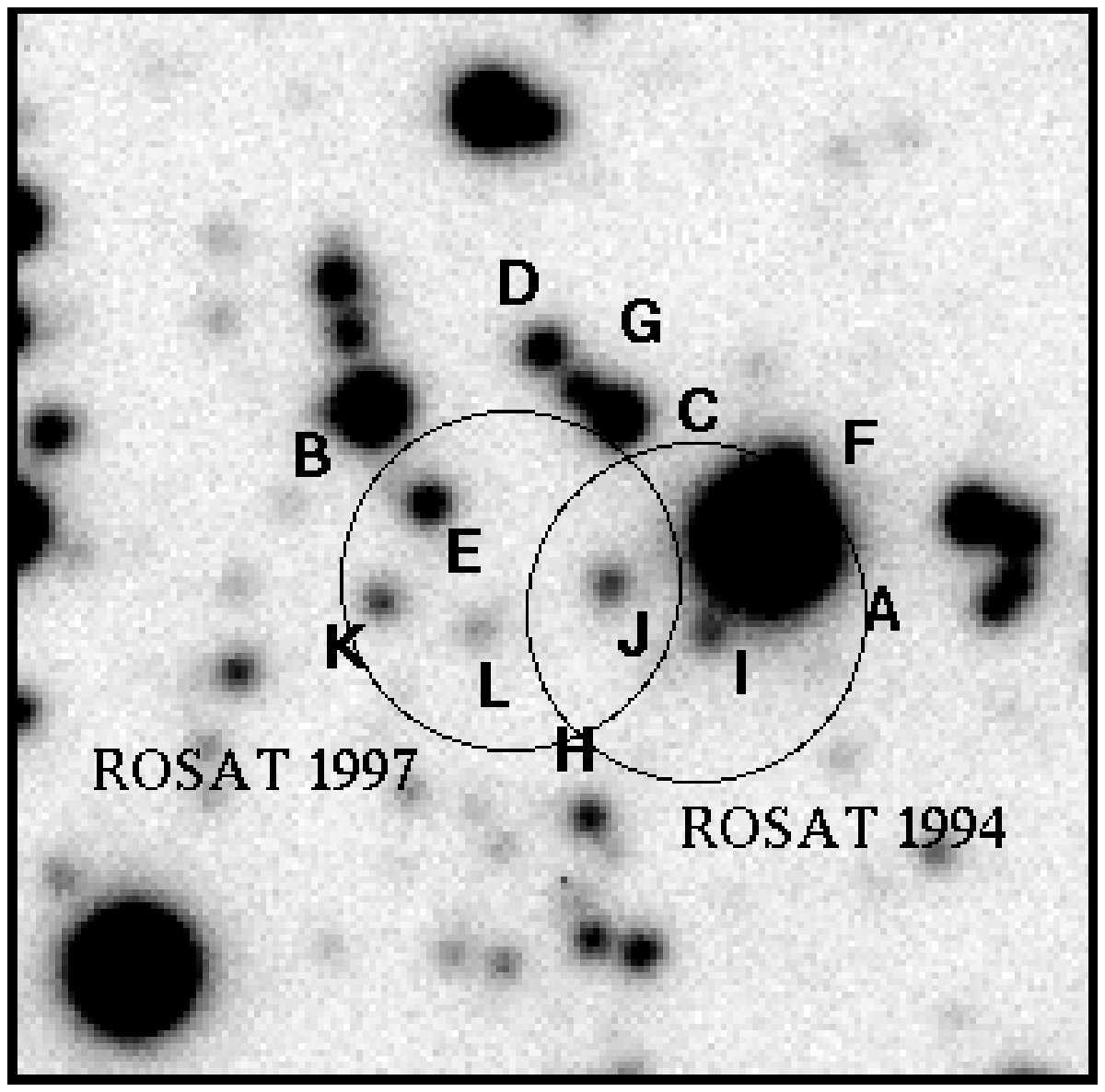}}
\bigskip
\caption{R image of the field of \src. 
The new X--ray position uncertainty regions (10\arcsec\ radius) obtained 
with the 1994 and 1997 HRI observations are also shown. 
North is top, east is left.}
\end{inlinefigure}

\section{Discussion}

The great majority of accreting X--ray pulsars are found in two 
distinct classes of massive binaries: systems with OB type supergiant
companions and systems with Be stars. The latter are often characterized
by transient activity.

We found that star A, the brightest candidate counterpart in the new 
error circles, is a main sequence B0 star, but the properties of its spectrum 
make an association with \src\ very unlikely.
The absence of emission--lines  corresponding to the Balmer 
series, HeI and FeII strongly suggest that star\,A is not in a binary 
system with a compact object. 
Although the optical emission--lines in Be/neutron star 
systems may occasionally disappear during the quiescence state,  
such a behaviour has been so far observed only in transient
sources which show also  outbursts and pronounced spin--up/down rate episodes. 
The low value of the inferred spin--down period derivative over a 6 month interval 
and the constant X--ray flux value over a timescale of 3 years (see below) 
rule against  this possibility. 
Furthermore, the soft X--ray spectrum of \src\ is much softer than 
that observed in all X--ray pulsars with massive companions
(i.e. a hard power law with photon index $\sim$1 that steepens only above
the cut--off energy of $\sim$10--30 keV). 
\begin{figure*}[!tb]
\psfig{figure=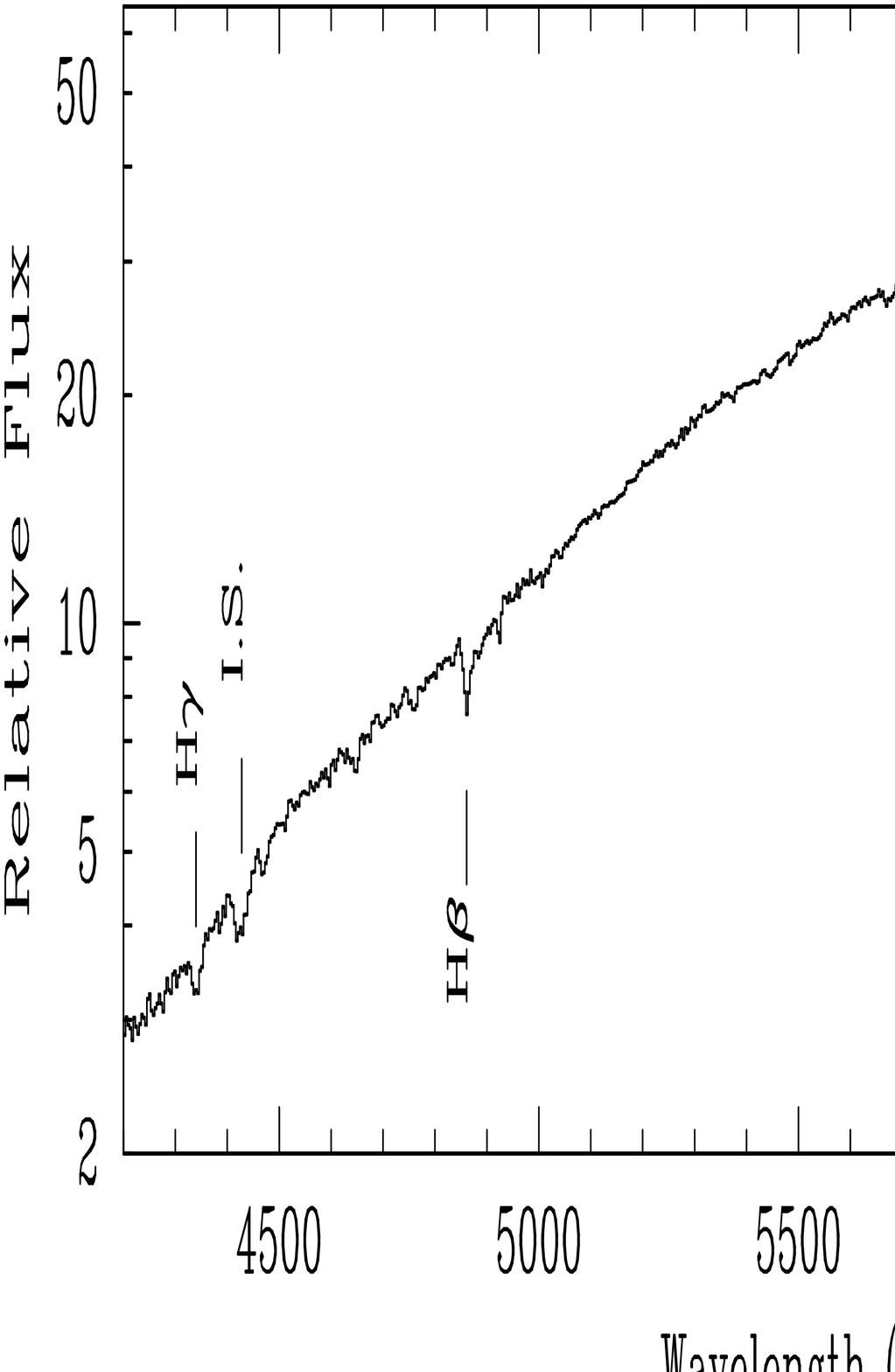,width=11cm,height=5cm} 
\caption{4300--6800 \AA\ low--resolution flux--calibrated spectrum of 
the reddened B0V spectral--type star A. I.S. stands for inter--stellar 
absorption.}
\end{figure*}

%\begin{table*}
\begin{inlinetable}
\begin{center}
\caption{Positions, R magnitudes and colors of the  stars within (or close to) 
the X--ray error circles of \src\ }
\begin{tabular}{ccccc}
\\ \hline
\hline
Star & R.A. (J2000)$^a$ & DEC (J2000)$^a$ & R$^b$ & V--R$^b$ \\
&       (hh mm ss)  &  ($\deg$ $\arcmin$ $\arcsec$) & mag  & mag \\
\hline
A & 17 08 46.2& --40 08 48& 13.41 & 1.48\\
B & 17 08 47.8& --40 08 40& 16.46 & 0.85\\
C & 17 08 46.8& --40 08 40& 17.36 & 1.48\\
D & 17 08 47.1& --40 08 36& 18.24 & 1.50\\
E & 17 08 47.6& --40 08 46& 18.26 & 1.27\\
F & 17 08 46.1& --40 08 43& 18.36 & 1.38\\
G & 17 08 47.0& --40 08 38& 18.60 & 1.50\\
H & 17 08 46.9& --40 09 05& 18.74 & 1.25\\
I & 17 08 46.4& --40 08 53& 18.86 & 1.61\\
J & 17 08 46.8& --40 08 50& 18.93 & 1.36\\
K & 17 08 47.8& --40 08 52& 19.00 & 1.38\\
L & 17 08 47.4& --40 08 53& 19.78 & 1.58\\
\hline
\end{tabular}
\end{center}
\begin{minipage}{0.87\linewidth}
\textsuperscript{a}--- Uncertainty $\sim$1\arcsec. \\
\textsuperscript{b}--- Absolute uncertainty $\sim$0.2 mag. 
Relative uncertainty $\sim$0.02 mag. 
\end{minipage}
%\noindent $^a$  \\
%\noindent $^b$ 
\end{inlinetable}
%\end{table*}

The second brightest candidate to be considered is  star B.
Its position in the CMD of Fig.~2 is consistent with a relatively closely 
K type star.

All remaining candidates are fainter than  R$\sim$17 implying
X--ray to optical flux ratios higher than those typically observed in 
High Mass X--ray Binaries. 
This is also indicated by their position in the CMD shown in Fig.~2:
the objects in the error circles 
are too blue and too faint to be compatible with  early type stars. 
Note also that our images are deep enough to exclude  very 
reddened and distant O and B stars. 
The absorption column density for \src\  is  
$\sim 1.5\times10^{22}$\,cm$^{-2}$ (Sugizaki et al. 1997), corresponding to
E(B--V) $\sim 2.6$ (Bohlin et al. 1978). Applying the standard
interstellar extinction law (Fitzpatrick 1998) this translates to
an absorption in the R band of A$_{\rm R}\sim 6$.
A B0 star located at a distance
of $5 \div 10$\,kpc would be observed with a R magnitude $\sim 15.5 \div
17$ and V--R $\sim 2$, i.e. it should have been easily detected in our 
optical images.

Our optical data  cannot exclude  that \src\ is part of a Low Mass X--ray Binary 
(LMXB). The faint stars in the error box   
are all compatible with relatively late spectral--type stars. 
For example star\,I  is consistent with a A5--7V at a distance of 3\,kpc, 
while star\,H 
is consistent with a late K or an early M at a distance of 
700\,pc. However, none of these objects has the characteristic blue 
color generally observed in the counterparts of luminous LMXBs,
in which the optical emission is dominated by the accretion disk (van Paradijs 1998).

The lack of a massive companion star and the inferred spin--down 
at a rate of $\sim$7.5$\times$10$^{-4}$ s yr$^{-1}$   strongly
support the inclusion of   \src\ in the group of AXPs, 
as was previously suggested by Sugizaki \etal (1997) 
based only on the soft X--ray spectrum and 11\,s period.

\src\ was detected by the \R\ HRI at the same 
flux level in both the 1994 and 1997 observations. 
The \R\ flux extrapolated to the 0.8--10 keV  range  
(2.3$\times$10$^{-11}$ erg cm$^{-2}$ s$^{-1}$; for the  power--law best fit) 
is about a factor of 2 lower than 
that observed with ASCA   (4.3$\times$10$^{-11}$ erg  cm$^{-2}$ s$^{-1}$).   
A soft blackbody component, accounting for 20--50\% of the total flux has been 
seen in the spectrum of three well studied AXPs (4U\,0142+61; White \etal 1996;  
Israel \etal 1999, 2E\,2259+587; Corbet \etal 1995;  Parmar \etal 1998, and 
1E\,1048.1--5937; Corbet \& Mihara 1997; Oosterbroek \etal 1998).
The presence of a similar component in \src\ was found 
compatible with the 
ASCA data (although not formally required by the fit,   Sugizaki \etal 1997).
Assuming a double component spectral model for   \src,  at least part of  the 
flux difference between the ASCA and \R\ observations might be  
ascribed to the more complex spectral slope than assumed. 
We conclude that the modest luminosity variations across observations
spanning three years is another analogy of \src\ with the AXPs class.
{\rc \src, like other AXPs, displays also a remarkable stability of the pulsed 
fraction and pulse shape.}
 
One of the main problems in understanding the nature of AXPs is to establish
whether they are  isolated neutron stars or members of binary systems
with very low mass companions.  Both deep optical and infrared
searches for counterparts and searches for Doppler orbital modulations
in the pulse frequency can provide valuable information in this respect.
In the best studied AXPs, the limits derived from such studies 
exclude most classes of companions stars, but helium stars 
with M$\leq 0.8 M_{\odot}$ and white dwarves (Mereghetti, Israel \& Stella 1998; 
Wilson \etal 1999).

The possibility  that \src\ has a white dwarf companion star is
compatible with our optical observations: the expected R magnitude 
at a distance of 700\,pc would be $\geq$ 21, well below the limiting magnitude of 
our images. Similar results apply to a helium main sequence star. We also note 
that although the absorption column inferred with the ASCA data would place 
\src\ quite far from us (10 kpc), it is possible that part of the absorption be
intrinsic.  
{\rc Also the possibility that \src\ is a
``magnetar'', an isolated strongly magnetic neutron star, possibly related to
Soft $\gamma$--ray Repeaters remains viable. More period measurements 
are clearly required to look for ``glitches'' as foreseen in the ``magnetar'' 
scenario.}
In conclusion, based on the  X--ray and optical findings presented 
here, we confirm that \src\ is a new member of the AXPs group.

\acknowledgments
GLI and SC thanks V.F. Polcaro for his kind help and useful discussions. 
This work was partially supported through ASI grants.


\vspace{3mm}
\begin{references}
\vspace{3mm}

\reference{} Bohlin, R.C., Savage, B.D., and Drake, J.F., 1978, \apj, 224, 132

\reference{} Campana, S., Lazzati, D., Panzera, R., et al., 1999, \apj, submitted

\reference{} Corbet, R.H.D., Smale, A.P., Ozaki, M., et al., 1995, \apj, 443, 786

\reference{} Corbet, R.H.D., and Mihara, T., 1997, \apjl, 475, L127

\reference{} Fitzpatrick, E.L., 1998, astro--ph/9809387

\reference{} Israel, G.L., Oosterbroek, T., Angelini, L., et al., 1999, \aap, in press 
(astro--ph/9904103)

\reference{} Lazzati, D., Campana, S., Rosati, P., et al., 1999, \apj, submitted 

\reference{} Kouveliotou, C., Dieters, S., Strohmayer, T., et al., 1998, \nat, 393, 235

\reference{} Mereghetti, S., 1999, in ``Relationship between Neutron Stars and 
 Supernova Remnants'', Mem.S.A.It., 
 in press (http://www.arcetri.astro.it/$\sim$elba98/index.html)

\reference{} Mereghetti, S., and Stella, L., 1995, \apjl, 442, L17

\reference{} Mereghetti, S., Israel, G.L.,  and Stella, L., 1998, MNRAS 296, 689

\reference{} Mereghetti, S., Stella, L., and Israel, G.L., 1998,  
 in ``The Active X--ray Sky: Results from BeppoSAX and RossiXTE'', Nuclear Physics B 
 Proceedings Supplements, Scarsi, L., Bradt H., Giommi, P., and Fiore, F. (eds), 
 Elsevier Science B.V. (Amsterdam), p. 253

\reference{} Oosterbroek, T., Parmar, A.N., Mereghetti, S.,  et al., 1998, \aap, 334, 925 

\reference{} Parmar, A.N., Oosterbroek, T., Favata, F., et al., 1998, \aap, 330, 175

\reference{} Stetson, P.B., 1987, \pasp, 99, 191

\reference{} Sugizaki, M., Nagase, F., Torii, K., et al., 1997, \pasj, 49, L25

\reference{} Taylor, J.H., and Cordes, J.M., 1993, \apj, 411, 674

\reference{} van Paradijs, J., Taam, R.E., and van den Heuvel, E.P.J., 1995, \aap, 299, L41

\reference{} van Paradijs, J., 1998, in ``The Many Faces of the Neutron Stars'',  
 Buccheri, R, van Paradijs, J., Alpar, M.A. (eds), Kluwer Academic Publishers (The 
 Netherlands), p.279

\reference{} Voges, W., Aschenbach, B., Boller, T.H., et al., 1998, 
 in ``New Horizons from Multi--Wavelength Sky Surveys'', 179th Symposium of the 
 International Astronomical Union, McLean, B.J., 
 Golombek, D.A., Hayes, J.J.E., and Payne, H.E. (eds), Kluwer Academic Publishers, p.33 

\reference{} White, N.E., Angelini, L., Ebisawa, K., et al., 1996, \apjl, 463, L83

\reference{} Wilson, C.A., Dieters, S., Finger, M., et al., 1999, \apj, 513, 464 


\end{references}
\end{document}